\documentclass[preprint]{aastex}
\usepackage{psfig}
\usepackage{amssymb}
\input{epsf.sty}
\def\eps@scaling{.95}
\def\plotone#1{\centering \leavevmode
   \epsfxsize=\eps@scaling\columnwidth \epsfbox{#1}}

\shorttitle{Deciphering the Milky Way}
\shortauthors{Gilmore, Wyse \& Norris}
\begin{document}
\title{Deciphering The Last Major Invasion of the Milky Way Galaxy}

\author{Gerard Gilmore}
\affil{Institute of Astronomy, Madingley Road, Cambridge CB3 0HA, UK}
\email{gil@ast.cam.ac.uk}

\author{Rosemary F.G.~Wyse\altaffilmark{1,2}}
\affil{The Johns Hopkins University, Dept.~of Physics and Astronomy,
Baltimore, MD 21218}
 \email{wyse@pha.jhu.edu}

\author{John E.~Norris}
\affil{Research School of Astronomy \& Astrophysics, The Australian
National  University, Mount Stromlo Observatory, Cotter Road, Weston,
ACT 2611} 
\email{jen@mso.anu.edu.au}

\altaffiltext{1}{School of Physics \& Astronomy, University of
St.~Andrews, Scotland, UK} 
\altaffiltext{2}{Astrophysics, Oxford University, England, UK}

\begin{abstract}

We present first results from a spectroscopic survey of $\sim 2000$
F/G stars 0.5--5kpc from the Galactic plane, obtained with the 2dF
facility on the AAT. These data show the mean rotation velocity of the
thick disk about the Galactic center a few kpc from the plane is very
different than expectation, being $\sim 100$km/s, rather than the
predicted $\sim 180$km/s.  
We propose that our sample is dominated by stars from a
disrupted satellite which merged with the disk of the Milky Way Galaxy
some 10-12Gyr ago.  We do not find evidence for the many 
substantial mergers expected in hierarchical clustering theories. We
find yet more evidence that the stellar halo retains kinematic
substructure, indicative of minor mergers.
\end{abstract}


\section{Introduction}

Mergers and strong interactions between galaxies happen, and may well
be the dominant process in the determination of a galaxy's current
Hubble type, particularly in the context of modern
hierarchical-clustering theories of structure formation (e.g.~Silk \&
Wyse 1993).  The recently discovered (Ibata, Gilmore \& Irwin 1994)
Sagittarius dwarf spheroidal galaxy is inside the Milky Way
Galaxy, is losing a significant stellar mass through tidal effects
(Ibata et al.~1997), forming star streams in the halo (Mateo, Morrison
\& Olszewski 1998; Ibata et al.~2001; Yanny et al.~2000), but having
little effect on the present structure of the bulk of the Galactic disk.

The outcome of a merger of two stellar systems depends on several
factors, most importantly the mass ratio and density contrast. During
a merger, energy, momentum and angular momentum are re-distributed so
that the common aftermath of a merger between a large disk galaxy and
a smaller, but still significant, satellite galaxy (more massive than
the Sagittarius dwarf spheroidal galaxy) is a heated disk and a
disrupted satellite (Quinn \& Goodman 1986; Velaquez \& White
1999). This is currently the most plausible model for the origin of
the thick disk in our Galaxy (see reviews in Gilmore, Wyse \& Kuijken
1989; Majewski 1993) and those of other galaxies; the
stochastic nature of the merger process allows for a wide variety of,
and indeed non-existence of, thick disks in external galaxies, as
observed, provided only a small number of merger events are involved.
Determination of the stellar populations in the Galactic thick disk
tests this model, and so constrains the merger history of the Milky
Way (Gilmore \& Wyse 1985; Wyse 2001).

All indications are that the Galactic thick disk is composed of only
very old stars, ages $\ga 10$~Gyr, equivalent to forming at a redshift
of $\ga 1$ (Wyse 2000).  This implies that the event that formed it from the thin
disk, which now contains stars of all ages, occurred a long time ago,
with little subsequent extraordinary heating of the thin disk.  If
this model is valid, it may be possible to identify stars captured
from the accreted galaxy, and to distinguish them from those formed in
the early thin disk of the Milky Way.  This would allow tight
constraints on what merged, and when it merged, and on the early star
formation in an extended disk.  These are important tests of
hierarchical clustering theories of structure formation.

\section{The Survey}

We are investigating the stellar populations of the Galactic thick
disk and halo through a statistical study of the kinematics (radial
velocities) and metallicity distributions of stars a few kiloparsecs
from the mid-Plane of the Galactic disk, down several lines-of-sight.
Our survey uses the two-degree-field multi-object spectrograph (2dF;
Lewis et al.~2002) on the Anglo-Australian Telescope, which provides
400 spectra simultaneously; we obtained data with spectral resolution
of 2.5\AA, in the wavelength range of 3700-4700\AA. The velocity
accuracy per star is $\sim$15 km/s, determined from repeat
observations and from a globular cluster standard.  Chemical abundance
determinations are in progress.  We here present our first kinematic
results, for around 2,000 stars.

Our survey is of F/G stars (${\rm B-V \leq 0.7}$) with V-band apparent
magnitudes in the range 17.0--19.5.  This preferentially selects main
sequence stars close to the turn-offs of the thick disk and halo
populations, at distances of several kiloparsecs.  Our primary fields
are at ($\ell, b$)= (270,--45) and (270,+33), against Galactic
rotation; thus radial velocities, in combination with a distance,
approximate Galactocentric orbital angular momentum, without the need
for transverse velocities. Bulge stars do not contribute at this
distance from the Galactic Center, while the apparent
magnitude/distance selection provides a strong bias against the thin
disk.

Star-count model predictions in one of our `rotation' fields for the
relative contributions of F/G stars belonging to each of the three
dominant populations of the Milky Way which contribute at the solar
neighborhood, as a function of apparent magnitude, are shown in
Figure~1, derived from a derivative of the star count model described
by Gilmore (1984).

\begin{figure}[ht!]
\psfig{figure=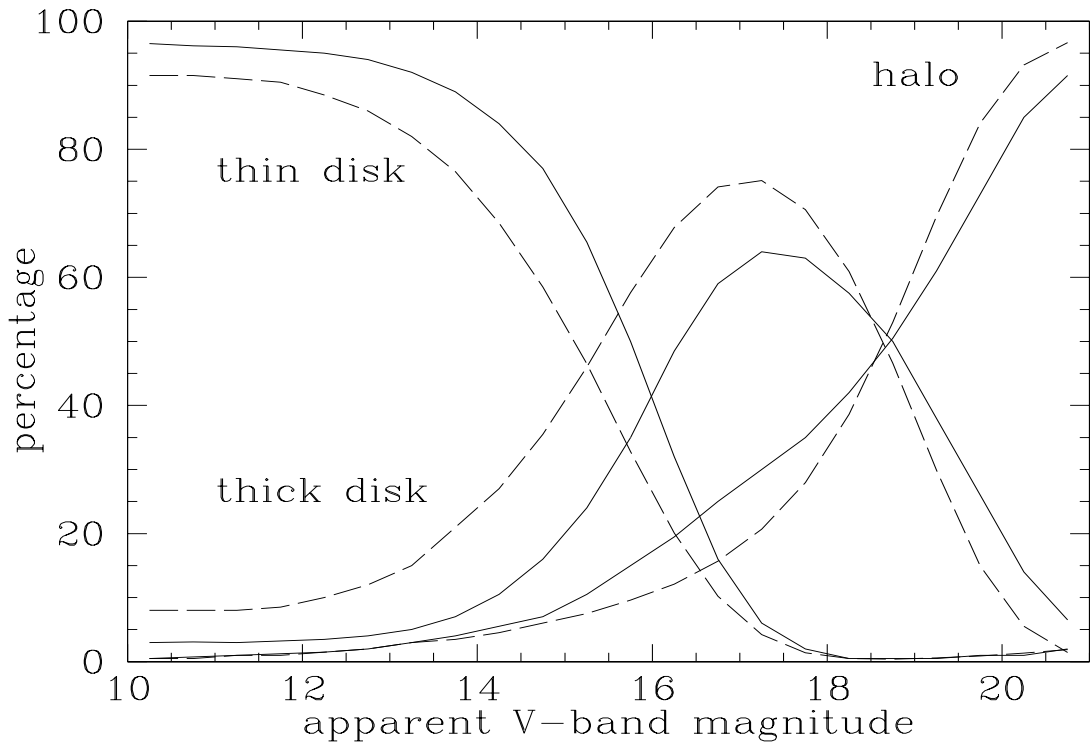,height=4in,width=6in,angle=270}
\caption{Star counts predictions for the relative contributions of the
thin disk, thick disk and stellar halo to the counts of unevolved F/G
stars in one of our fields (they do not differ substantially between
our fields).  The predictions with two different normalisations and
scale-heights for the thick disk are shown, to illustrate the current
range of plausibility. The first model (solid curves) has local
normalisation of 2\% and scaleheight of 1300~pc and the second (dashed
curves) has local normalisation of 8\% and scaleheight of
850~pc. Standard parameters for thin disk and stellar halo are
assumed. }
\end{figure}

\section{Results}

We detect a substantial population of stars on orbits that are
intermediate between those of the canonical thick disk and the
canonical stellar halo.  As metallicities are not yet available, we
sort on distance statistically, using apparent magnitude, since all
stars observed have similar colours. Figure~2 shows the line-of-sight
(radial) velocity histograms for stars which have spectra with
signal-to-noise greater than 20, in the lines-of-sight that probe
orbital angular momentum.  The upper panel shows the distributions for
the statistically nearer sub-sample, stars with $V < 18$, while the
lower panel shows more distant stars, $18 \leq V < 19.5$ (we have
observed twice as many 2dF pointings of faint stars in the
line-of-sight at $b = +30$, resulting in approximately twice as many
radial velocities in this field, compared to the $b=-45$ field).
These data are compared with the predictions of a model with thin
disk, thick disk and halo with standard stellar kinematics, resulting
from the median of 11 random samplings of three Gaussians with means
and dispersions given in Table~1 (based on Dehnen \& Binney 1998 and
Chiba \& Beers 2000). The model kinematics were fixed, independent of
magnitude range, and the relative proportions of each component are
chosen within the ranges in Figure~1, for the characteristic magnitude
of each sample, namely $V = 17.5$ for the upper panel with
thin:thick:halo of 5:69:26, and $V=18.5$ for the lower panel with
thin:thick:halo of 0.5:54.5:45.  Using the Bergbusch \& vandenBerg
(2001) isochrones, at these magnitudes typical thick disk stars with
[Fe/H] $\sim -0.5$~dex and ${\rm B-V = 0.6, \, M_V \sim +4.9}$ would
be at distances from the Sun of $\sim 3.3$~kpc and $\sim 5.3$~kpc,
while typical halo stars with metallicity of $\sim -1.5$~dex and ${\rm
B-V = 0.5, \, M_V \sim +5.4}$ would be at distances of $\sim 2.6$~kpc
and $\sim 4.2$~kpc respectively.

The data for the brighter stars and the model are in tolerable
agreement showing the well-established canonical thick disk lag
of less than 50km/s, albeit that there is a noticeable difference
between the radial velocity distributions in the two fields. However,
there is an obvious disagreement in the lower panel, for the fainter
stars, in that the typical star shows a mean lag behind the Sun of
$\sim 100$km/s.  The peak of the observed distribution is
significantly displaced from the model predictions.  This disagreement
is not sensitive to the adopted normalisations or scale heights for
the thick disk and halo, but indicates the need for a substantial
revision in the standard kinematical model of the Milky Way.

We emphasise that the number-magnitude-colour distribution of stars
seen in our fields is similar in both fields (many kiloparsecs apart
at these apparent magnitudes) and is consistent with the predictions
of standard star count models. We have detected not a small perturbation
superimposed on a smooth well-understood
background, but rather intrinsic complexity in the kinematic
distribution function of stars ascribed in standard models to the
thick disk.  We also note that the predictions of the Gaussian halo
fail to reproduce the local peak in the data at around 300~km/s, which
is suggestive of a retrograde halo stream (velocities above $\sim
180$~km/s in these lines-of-sight are retrograde), as may be produced
by accretion of a small satellite (e.g.~Helmi et al.~1999). This
feature will be discussed elsewhere.

\begin{figure}[ht!]
\psfig{figure=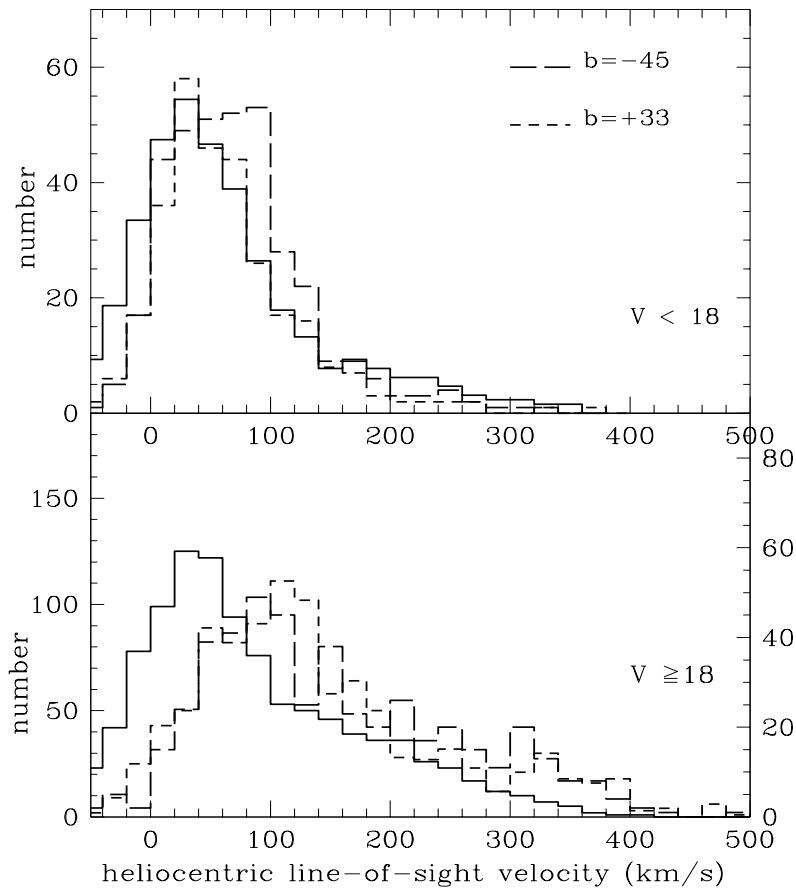,height=5in,width=4.1in}
\caption{The radial velocity histograms for F/G stars in two
lines-of-sight, compared to model predictions.  The kinematics of the
brighter stars, with apparent magnitude less than $18$ in the V-band,
are shown in the top panel and the kinematics of the fainter stars are
shown in the bottom panel.  The solid histograms result from random
sampling Gaussians with `standard' kinematics, as described in the
text.  The short-dashed histograms are the data for the field at
$b=+33$, in which at these distances the heliocentric line-of-sight
velocity corresponds to $\sim 80$\% of rotation velocity, while the
long-dashed histograms are the data for the field at $b = -45$ in
which it corresponds to $\sim 70$\%.  In the lower panel the y-axis
scale on the right-hand-side refers to the long-dashed histogram.  }
\end{figure}

Figure~3 compares our results for the more distant stars with the
predictions of a revised model, where the adopted kinematics of the
`thick disk' stars are instead those of the `satellite debris' as
given in Table~1, all other quantities being held fixed.  The
agreement with the velocity distribution is much improved, except (as
expected) for the local maximum at V$\ga$300~km/s.

\begin{figure}[ht!]
\psfig{figure=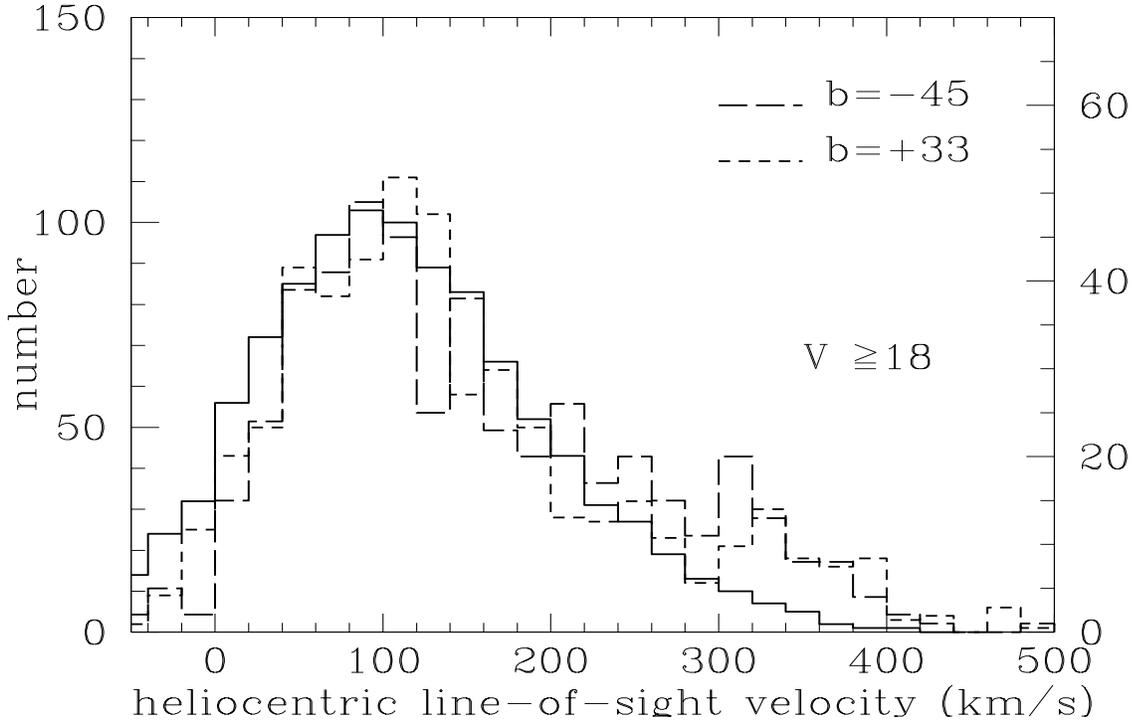,height=4in,width=6in,angle=270}
\caption{As the lower panel of Fig.~2, but now with the model modified
such that the kinematics of the `satellite debris' from Table~1 has
been adopted for the `thick disk' stars.  Much improved agreement is
seen, except for V$\ga$300~km/s.  }
\end{figure}

Our results demonstrate that far from the Galactic plane the `thick
disk' is a population which has significantly lower mean rotational
velocity, and most probably higher values of velocity dispersion, than
it has close to the midplane. Previous investigations have seen
indications of this `vertical shear' at lower Z-distances (Wyse \&
Gilmore 1986; Majewski 1993; Chiba \& Beers 2000; and indeed one can
see indications in the upper panel of Fig.~2, especially the field at
$b=-45$), but the observations had been interpreted in terms of a
smooth gradient with Z-height within a population characterised by one
vertical velocity dispersion.  Thus to explain the large change in
mean azimuthal streaming seen in the observations would require a disk
distribution function such that the angular velocity of stars on the
orbits sampled by observing 2-3 kpc vertically from the disk, at
approximately fixed radial distance, differs by a factor of two from
those samples at lower Z-heights.  An effect of this large an
amplitude is hard to envisage.  A more straightforward interpretation
is that the `thick disk' has a distribution function which has a
composite functional form; whether this is actually a continuum or
superposition of discrete components remains to be seen.

\section{Implications}

This new appreciation that a significant fraction of `thick disk
stars' are on orbits with angular momentum around half that of the
Sun's orbital angular momentum resolves a recently discovered puzzle
in the white dwarf population of the Galaxy.  Two recent surveys
(Oppenheimer et al.~2000; Nelson et al.~2002) have detected candidate
white dwarfs, based on their high transverse motion across the sky, in
numbers that are significantly higher than expected on the basis of
normal thin- and thick-disk kinematics.  This has led to speculations
of a baryonic dark halo (Oppenheimer et al.~2000; Nelson et al.~2002),
However, our result that typical stars on orbits that take them far
from the Plane have higher than expected mean velocities increases the
expected detections in any proper-motion-selected sample, and resolves
the discrepancies (Gilmore et al., in prep).  Indeed one expects
`shredded satellite' stars to be found in local,
proper-motion-selected samples.  Just such a population, with a mean
azimuthal stream velocity of $\sim 100$~km/s, was identified by Fuchs,
Jarheiss \& Weilen (1999), based on the subset of the Carney et
al.~(1994) sample with Hipparcos distances and proper motions
(resulting in only a handful of stars in the `excess' population).

What do our results mean for the evolution of our Galaxy, presumably a
typical disk galaxy? In the standard hierarchical clustering and
merging picture of galaxy formation a thick disk is an expected
outcome of a significant merger. Depending on the mass, density
profile and orbit of the merging satellite, `shredded-satellite' stars
may retain a kinematic signature distinct from that part of the thick
disk that results from the heated thin disk.  Satellites on prograde
(rather than retrograde) orbits couple to the rotating
thin disk more efficiently, and thus a merger with such a system is
favored as the mechanism to form the thick disk (Quinn \& Goodman
1986; Velazquez \& White 1999).  If any kinematic trace of the
now-destroyed satellite galaxy is visible, it will be seen in the mean
orbital rotational velocity of stars.  The actual lag expected from
the shredded-satellite depends predominantly on the initial orbit and
the amount of angular momentum transport in the merger process, and is
not {\sl ab initio} predictable in a specific case.

In general however, excluding special initial conditions, such as a
circular orbit at large distance (see Walker, Mihos \& Hernquist
1996), satellite debris stars will be on orbits characterised by lower
net rotational streaming about the Galactic Center than that of the typical 
scattered former thin-disk star at a given distance from the Galactic
center.  In order to support themselves against the Galactic potential
with less angular momentum support, the shredded-satellite debris must
then have larger random motions (equivalent to pressure) than do the
typical thick disk stars: this is seen in numerical simulations of
this process (Walker et al.~1996). It is these kinematics which allow
their detection: stars with the highest amplitude of vertical motions
(the satellite debris?) will be preferentially found farther from the
Plane than are most thick disk stars (the heated thin disk?).  If a
population of former satellite stars exists, and the satellite was on
an initial non-circular orbit consistent with cosmological simulations
(van den Bosch et al.~1998), the apparent mean rotational velocity of
stars far from the Plane (the debris) will be less than it is for
stars near the Sun, in the `classical' thick disk.  This situation is
easily distinguished from the possibility that all stars form a
single, coherent, thick disk, in which case the rotational velocities
of the most distant thick disk stars, far from the Plane, will not
differ significantly from those nearby.  This second model is
inconsistent with our observations.

Our observations favor the interpretation that we have detected the
shredded satellite.  This is then evidence for a significant past
merger experienced by the Milky Way Galaxy, in the form of the direct
detection of the intruder stars, stars that were formed in another
galaxy, long ago.  Decomposition of the star counts into `heated thin
disk' and `shredded satellite', on the basis of the kinematics,
suggests that a galaxy with perhaps one-quarter of the stellar mass of
the early thin disk was the cause of the formation of the thick disk. 
The satellite debris may account for some of the distant 
structures seen by Newberg et al.~(2002), but more data are needed to establish this. 
We do not see several distinct `extra' peaks in the velocity
histograms, implying, in this interpretation, only one major past
merger. 

Alternative explanations remain viable.  We may have found the remnant
of a early merger event which occurred prior to a later merger which
formed the bulk of the present thick disk. Outside the standard
hierarchical merger-based cosmology, models without mergers exist: a
vertically extended disk could have formed during the initial cooling
and contraction of proto-disk gas to reach equilibrium in a thin
configuration (Norris \& Ryan 1991; Burkert, Truran \& Hensler 1992).
In this case one then expects smooth vertical gradients in kinematics and
metallicity within the continuous disk, now envisaged to encompass
all of what we have identified as the thin disk, the thick disk, and the
`shredded satellite'.

The bivariate distribution function of kinematics and chemical
abundances provides more information than just kinematics: abundance
determinations for our sample are underway.  Smooth settling scenarios
make specific predictions, but early merger models depend on the {\sl
a priori\/} unknown properties, at high redshift, of a now-destroyed galaxy.
It is unlikely, however, that the chemical abundance
distribution of the shredded satellite will be similar to that of the
heated thin disk. Future chemical element ratio studies will be able
to limit the star formation histories of thick disk stars as a
function of vertical velocity dispersion, and quantify many of these
general statements.

We thank Ken Freeman for his contributions to this
project.  RFGW acknowledges support in the UK from a PPARC Visitor
Grant, and thanks all at Oxford and St Andrews for stimulating
environments.

%%%%%%%%%%%%%%%%%%%%%%%%%%%%%%%%%%%%%%%%%%%%%%%%%%%%%%%%
%% Tables %%
%%%%%%%%%%%%%%%%%%%%%%%%%%%%%%%%%%%%%%%%%%%%%%%%%%%%%%%%
%\clearpage

\begin{deluxetable}{lcccc}
%\footnotesize
\normalsize
\tablecaption{\large {Stellar Kinematics }}
%\label{data.tab}}
\tablewidth{0pt}
\tablehead{
\colhead{Component} & \colhead{$\sigma_W$ } & \colhead{$< V_{\rm lag}>$} & \colhead{$\sigma_V $} & \colhead{Z$_{\rm max}$ (pc) }  \\
\colhead{} & \colhead{} & \colhead{} & \colhead{} & \colhead{ for $W=\sigma_W$ } 
}
\startdata
old thin disk  & 20 & 20  & 25 &  300  \\ 
local thick disk  & 35 & 35 & 50 & 600   \\
stellar halo & 95 & 220 & 105 & 3000   \\
satellite debris?  & 60  & 100  & 70  & 1200   \\
\enddata
\tablecomments{Z$_{\rm max}$ calculated using the vertical
potential at the solar circle derived by Kuijken \& Gilmore (1989). }

\end{deluxetable}

\end{document}